\documentclass{aa}

\usepackage{graphicx}
\usepackage{txfonts}
\newcommand{\rhoO}{$\rho$~Ophiuchi }

\begin{document}

   \title{The Gaia-ESO Survey: \\Dynamical Analysis of the L1688 region in Ophiuchus}

   \author{E. Rigliaco\inst{1}, B. Wilking\inst{2}, M.R. Meyer\inst{1}, R. D. Jeffries\inst{3}, M. Cottaar\inst{1}, A. Frasca\inst{4}, N.~J. Wright\inst{5}, A. Bayo\inst{6}, R. Bonito\inst{7,8}, F. Damiani\inst{8}, R.~J. Jackson\inst{3}, F. Jim\'enez-Esteban\inst{9,10}, V. M. Kalari\inst{11}, A. Klutsch\inst{4}, A.~C. Lanzafame\inst{12,4}, G. Sacco\inst{13}, G. Gilmore\inst{14}, S. Randich\inst{13},  E.~J. Alfaro\inst{15}, A. Bragaglia\inst{16}, M.~T. Costado\inst{15}, E. Franciosini\inst{13}, C. Lardo\inst{17}, L. Monaco\inst{18}, L. Morbidelli\inst{13}, L. Prisinzano\inst{8}, S.~G. Sousa\inst{19}, S. Zaggia\inst{20}
          }

\institute{Institute for Astronomy, Department of Physics, ETH Z{\"u}rich, Wolfgang-Pauli-Strasse 27, 8046 Z{\"u}rich, Switzerland \\
              \email{elisabetta.rigliaco@phys.ethz.ch}
\and   
Department of Physics and Astronomy, University of Missouri—St. Louis, 1 University Boulevard, St. Louis, MO 63121, USA;
\and
Astrophysics Group, Research Institute for the Environment, Physical Sciences and Applied Mathematics, Keele University, Keele, Staffordshire ST5 5BG, United Kingdom
\and
INAF - Osservatorio Astrofisico di Catania, via S. Sofia 78, 95123, Catania, Italy
\and
Centre for Astrophysics Research, University of Hertfordshire, Hatfield AL10 9AB, United Kingdom
\and
Instituto de F\'isica y Astronomi\'ia, Universidad de Valpara\'iso, Gran Bretana 1111, Playa Ancha, Chile
\and
Dipartimento di Fisica e Chimica, Universit\'{a} di Palermo, Piazza del Parlamento 1, 90134, Palermo, Italy
\and 
INAF - Osservatorio Astronomico di Palermo, Piazza del Parlamento 1, 90134, Palermo, Italy
\and
Centro de Astrobiolog\'{\i}a (INTA-CSIC), Departamento de  Astrof\'{\i}sica, PO Box 78, E-28691, Villanueva de la Ca\~nada, Madrid, Spain
\and
Suffolk University, Madrid Campus, C/ Valle de la Viña 3, 28003, Madrid, Spain
\and
Armagh Observatory, College Hill, Armagh BT61 9DG, United Kingdom
\and
Dipartimento di Fisica e Astronomia, Sezione Astrofisica, Universit\'{a} di Catania, via S. Sofia 78, 95123, Catania, Italy
\and
INAF--Osservatorio Astrofisico di Arcetri, Largo E. Fermi, 5, I-50125 Firenze, Italy
\and
Institute of Astronomy, University of Cambridge, Madingley Road, Cambridge CB3 0HA, United Kingdom
\and
Instituto de Astrof\'{i}sica de Andaluc\'{i}a-CSIC, Apdo. 3004, 18080, Granada, Spain
\and 
INAF - Osservatorio Astronomico di Bologna, via Ranzani 1, 40127, Bologna, Italy
\and
Astrophysics Research Institute, Liverpool John Moores University, 146 Brownlow Hill, Liverpool L3 5RF, United Kingdom
\and
Departamento de Ciencias Fisicas, Universidad Andres Bello, Republica 220, Santiago, Chile
\and
Instituto de Astrof\'isica e Ci\^encias do Espa\c{c}o, Universidade do Porto, CAUP, Rua das Estrelas, 4150-762 Porto, Portugal
\and 
INAF - Padova Observatory, Vicolo dell'Osservatorio 5, 35122 Padova, Italy
        }

   \date{Received August 26 2015; accepted January 09 2016}

 
  \abstract{The Gaia ESO Public Spectroscopic Survey (GES) is providing the astronomical community with high-precision measurements of many stellar parameters including radial velocities (RVs) of stars belonging to several young clusters and star-forming regions. One of the main goals of the young cluster observations is to study of their dynamical evolution and provide insight into their future, revealing if they will eventually disperse to populate the field, rather than evolve into bound open clusters.}  { In this paper we report the analysis of the dynamical state of  L1688 in the $\rho$~Ophiuchi molecular cloud using the dataset provided by the GES consortium. }
{We performed the membership selection of the more than 300 objects observed. Using the presence of the lithium absorption and the location in the Hertzspung-Russell diagram, we identify 45 already known members and two new association members. We provide accurate RVs for all 47 confirmed members.} {A dynamical analysis, after accounting for unresolved binaries and errors, shows that the stellar surface population of L1688 has a velocity dispersion $\sigma \sim$1.14$\pm$0.35~km~s$^{-1}$ that is consistent with being in virial equilibrium and is bound with a $\sim$80\% probability.  We also find a velocity gradient in the stellar surface population of $\sim$1.0~km~s$^{-1}$pc$^{-1}$ in the northwest/southeast direction, which is consistent with that found for the pre-stellar dense cores, and we discuss the possibility of sequential and triggered star formation in L1688. 
}{}
\authorrunning{E. Rigliaco et al.}
\titlerunning{The Gaia-ESO Survey: Dynamical Analysis of the L1688 Region in Ophiuchus}
   \maketitle
%

\section{Introduction}

The majority of stars form in groups or clusters inside molecular clouds. After 5--10~Myr, 90\% of the embedded clusters do not evolve to become bound open clusters, such as the Pleiades, but rather disperse into the field (Lada \& Lada 2003). 
This happens either due to the cluster's formation in an unbound state or becoming unbound during a $\sim$10~Myr timeframe from the dynamical evolution of stars within the cluster (Clark et al. 2005; Carpenter 2000;  Adams \& Myers 2001;  Lada \& Lada 2003), the expulsion of residual gas left over from star formation (e.g., Hills 1980; Lada et al. 1984; Goodwin \& Bastian 2006), the tidal heating from nearby Giant Molecular Clouds (Elmegreen \& Elmegreen 2001, Kruijssen 2014), or the photoionizing radiation from O stars in more or less massive clusters (Dale \& Bonnell  2011; Walch et al. 2012;  Dale et al. 2015). 
The fate of the cluster also affects planet formation because of frequent stellar encounters in crowded regions (e.g., Adams \& Laughlin 2001, Parker \& Quanz 2012), stellar multiplicity through dynamical interactions and the orbital separation distribution of binary systems (Parker \& Meyer 2014), and mass segregation towards the cluster core (Parker \& Reggiani 2013).   
A detailed study of clusters' dynamical states in a variety of environments is needed in order to understand the evolution of stellar clusters and the relative importance of these processes.

In recent years, many efforts have been made to determine precise radial velocities and studies of the dynamical states of young clusters are gaining new momentum.  
In this paper, we focus on results obtained with the Gaia-ESO large spectroscopic survey (GES). GES is providing astronomers with high-resolution optical spectra of stars in star-forming regions and clusters, and in the halo, bulge, and thick and thin disk of the Milky Way (Gilmore et al. 2012; Randich et al. 2013).  The GES data are acquired with the FLAMES multi-object spectrograph mounted on the Very Large Telescope, with both the GIRAFFE and UVES spectrographs. 
One of the main aims of the young cluster observations is to study their kinematics and dynamical evolution through the measurement of accurate radial velocities (Lanzafame et al. 2015).  The high-quality of GIRAFFE and UVES spectra also allows us to study their metallicities and elemental abundances, rotational velocities, chromospheric activity, and accretion rates (e.g., Spina et al. 2014; Frasca et al. 2015). 
So far, the dynamical properties of several young clusters have already been observed and analyzed.  
Using GES data, Jeffries et al. (2014) analyzed $\gamma$~Velorum, a $\sim$10--20~Myr old cluster.  They found two different kinematic populations: one population has an intrinsic dispersion of 0.34$\pm$0.16~km~s$^{-1}$ that is consistent with virial equilibrium and a second one with an intrinsic dispersion of 1.60$\pm$0.37~km~s$^{-1}$  that is composed of a scattered population of unbound stars. 
Sacco et al. (in prep.) is studying  the dynamics of the $\sim$2~Myr old Chamaeleon {\rm I} star-forming region, finding a velocity dispersion of $\sim$1.02$\pm$0.14~km~s$^{-1}$. 

Other radial velocity surveys are exploring this field.  
Foster et al. (2015) have analyzed the dynamical state of the young (1--2 Myr) cluster NGC~1333 using measurements of the RV carried out with the APOGEE (Apache Point Observatory Galactic Evolution Experiment) infrared spectrograph (Wilson et al. 2012). 
They have found that the velocity dispersion of $\sim$0.92$\pm$0.12 km~s$^{-1}$ is consistent with the virial velocity of the region and the diffuse gas velocity dispersion.  
Using APOGEE data, Cottaar et al. (2015) have analyzed the dynamical state of the 2--6~Myr old cluster IC~348. The velocity dispersion of 0.72$\pm$0.07~km~s$^{-1}$ implies a super-virial dynamical state.  
More massive clusters such as Westerlund {\rm I},  NGC~3603, and R136 have been studied by Cottaar et al. (2012), Rochau et al. (2010), and H\'enault-Brunet et al. (2012), respectively. Kinematic studies for the Orion Nebula Cluster (ONC) and NGC~2264 also exist (Furesz et al. 2006; Furesz et al. 2008; Tobin et al. 2015), together with studies of Cyg~OB2 (Wright et al. 2014, Parker et al. 2014). 
The analysis of the dynamical states of several clusters within their first Myr shows that they can be either bound or unbound, depending on their properties such as the star and gas density, star formation efficiency, and gas expulsion timescale (see e.g., Baumgardt \& Kroupa 2007). The role and importance of different environmental conditions and physical processes in determining the dynamical state of the cluster (bound or unbound) as a function of its age will be made clearer at the end of the GES, when observations for $\sim$30 young ($<$100~Myr) clusters will be completed.
 
In this paper we present an analysis of the dynamical state of the L1688 star-forming region in the $\rho$ Ophiuchi molecular cloud complex. We refer to Wilking et al. (2008) for a complete review of this well known star-forming region. In summary, it is located at $\sim$135~pc (Mamajek 2008) with over 300  members and a surface population median age between 2--5~Myr.   
The main cloud is L1688 with its $\sim$1$\times$2~pc centrally concentrated core. It has been the focus of numerous surveys in the near-infrared (e.g., Greene \& Young 1992, Cutri et al. 2003), mid- to far-infrared (e.g., Bontemps et al. 2001; Evans et al. 2005), X-ray (e.g., Gagn\'e et al. 2004; Ozawa et al. 2005), and submillimeter/millimeter continuum (e.g., Andr\'e \& Montmerle 1994; Pattle et al. 2015). Analysis of the proper motions in the cluster has also been recently conducted (Wilking et al. 2015). 
Although it is one of the closest regions of active star formation, optical surveys of this region are not numerous because of the high visual extinction in the cloud core (A$_V\sim$50--100~mag, e.g., Wilking \& Lada 1983). The most complete extinction-limited optical spectroscopic survey of young stellar objects (YSOs) in L1688 identified 135 candidate members (Wilking et al. 2005, hereafter WMR05; Erickson et al. 2011, hereafter E11). 
However, dynamical studies of this surface population have not been conducted because of the lack of precise RV measurements. 
The analysis of pre-stellar cores within L1688, conducted using the N$_2$H$^+$ molecule, has shown that they are either bound or virialized (Pattle et al. 2015) with a sub-virial velocity dispersion of $\sim$0.4~km~s$^{-1}$ (Andr\'e et al. 2007). 

We have conducted the GES observations in the direction of L1688 to analyze the radial velocity distribution of the stellar surface population, and to compare the latter to the pre-stellar core velocity dispersion.  In Sect. 2 we summarize the Gaia-ESO observations in the direction of L1688 and the data reduction and data analysis procedures. In Sect. 3 we describe the candidate member selection and in Sect. 4 we present the analysis of the radial velocity measurements and discuss the dynamical state of the star-forming region. We outline the main findings arising from our analysis in Sect.5. 


\section{Gaia-ESO survey: L1688}

This work is based on the results of the analysis of the spectra observed with GIRAFFE and UVES. The products have been released to the Gaia-ESO consortium as internal data release iDR2iDR3. Moreover, the GES consortium re-analyzed archival data collected from the ESO Archive which we add to our analysis. 

\subsection{Target selection \& observations} 

The Gaia-ESO target selection in the direction of L1688 is based on the photometric survey of Wilking et al. (1997, complemented with unpublished data) and the location of the targets in the $(R-I)$ vs. I color-magnitude diagram. The selection criteria for all the clusters analyzed within the GES are summarized in Bragaglia et al. (in prep.). 
In L1688, the targets have been selected among the stars with the following selection criteria: 
{\it i)} they have an estimated I-band magnitude I$\lesssim$17.5 mag for GIRAFFE targets, and I$\lesssim$14.0 mag for UVES targets and 
{\it ii)} they are located within the L1688 star-forming cloud in Ophiuchus (i.e.,
$16^h 30^m < $RA$ < 16^h 24^m$, $-25^\circ 12^\prime < $DEC $< -23^\circ 48^\prime$, Ridge et al. 2006). 

\begin{figure}
   \centering
    \includegraphics[angle=-90,width=9cm]{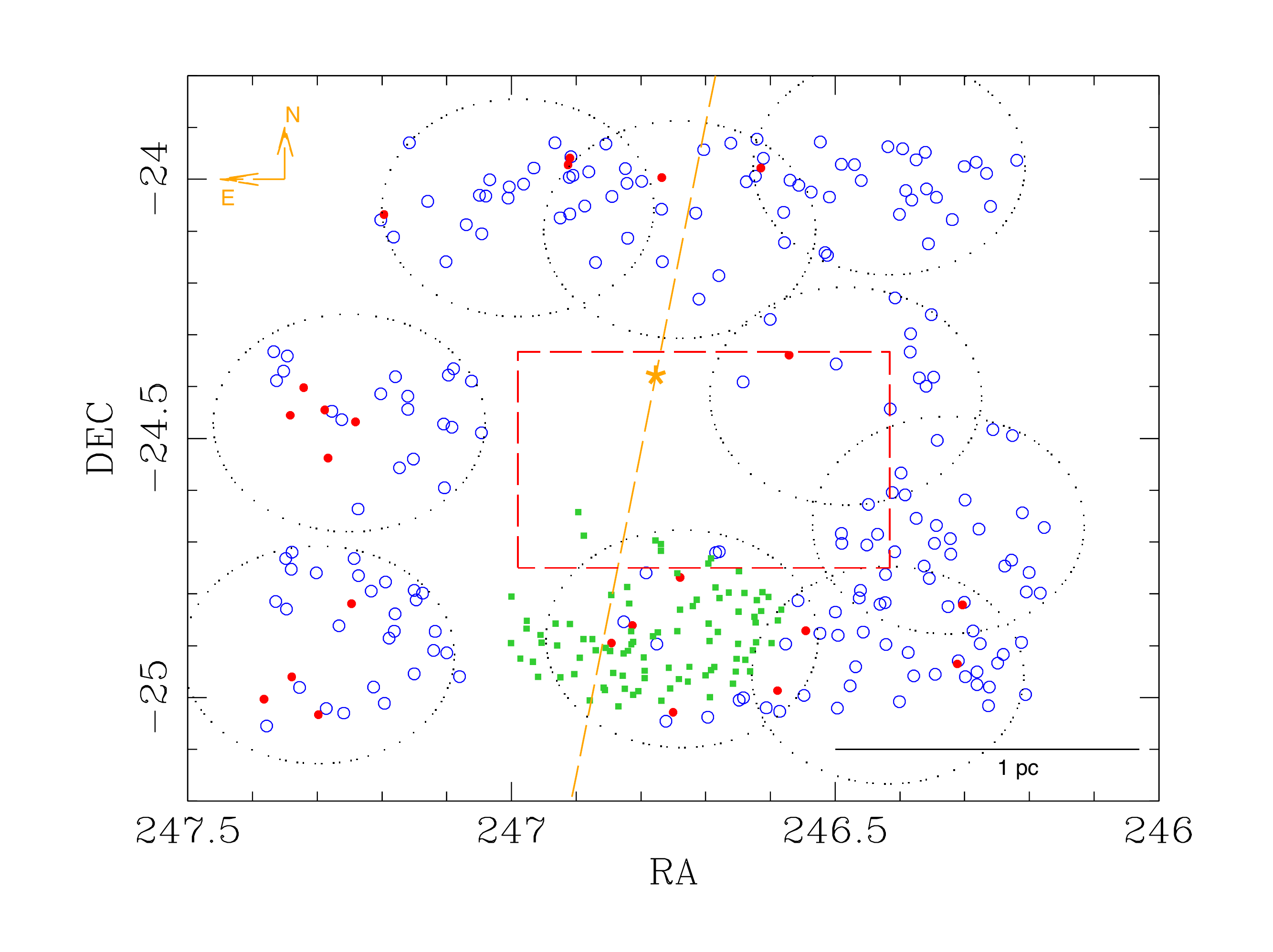}
   \caption{Map of the nine FLAMES fields of view observed with GES in the
direction of L1688: blue open circles show the GIRAFFE targets, red dots
represent the UVES targets, and green filled squares represent the GIRAFFE data
collected from the ESO archive. Dotted lines outline 25 arcmin diameter fields of view. The orange star shows the median location of the star positions and the orange-dashed line represents the northwest/southeast direction that is introduced in Sect. 4.3. The dashed box marks the location of the dense molecular cloud containing the pre-stellar cores. }
              \label{FigFOV}%
    \end{figure}

The observations were performed with the FLAMES fiber-fed
spectrograph in Medusa feeding mode, allowing the simultaneous allocation
of UVES high-resolution (R=47,000) and GIRAFFE intermediate resolution  (R = 17,000) 
fibers. The Medusa system allows for the allocation of 132 fibers per pointing with GIRAFFE, including the 
sky and 8 simultaneous UVES fibers. 
The fiber allocation performed during the 
L1688 observations never reached more than a few tens of fibers assigned due to 
the crowded environment, the limitation of 11 arcsec of separation between allocated fibers to avoid fiber collisions, and 
other instrumental limitations (e.g., guide star selection, Pasquini et al. 2002). 
The log of the observations, carried out on the
nights of 22-24-25 June 2012, and 30 August 2012, is reported
in Table~\ref{TabLOG}, along with central position and number of fibers allocated during  
each pointing on GIRAFFE, UVES, or sky targets. Due to the overlap between fields of view we have
observed 30 objects twice. As shown in Fig.~\ref{FigFOV}, nine FLAMES
pointings have been used to cover the region around L1688, avoiding
the central region of the cluster, where the extinction is very
high and the density of visible sources very low (A$_V$=50--100~mag, Wilking \& Lada 1983). The GIRAFFE observations
were performed using the HR15N setup covering the
wavelength range 6470–6790\AA, while the Red 580
setup (centered at 5800\AA) was used to acquire UVES
spectra. Altogether, 200 GIRAFFE targets and 23 UVES targets
were observed for a total of 223 objects.

We include in our analysis data contained in the ESO Archive for which FLAMES observations were available in the same region. In total, 90 objects observed
with GIRAFFE with the same setup as the GES data have been
re-reduced and re-analyzed by the GES team (green squares in
Fig.~\ref{FigFOV}). These objects belong to the 075.C-0256 ESO program
(PI Pallavicini) and were observed in 2005.

Considering GES and archival data, we thus conduct our analysis on a total sample of 313 objects.

%
\begin{table}
\caption{Log of the FLAMES pointings in the direction of L1688}             
\label{TabLOG}      
\centering                          
\begin{tabular}{c c c c}        
\hline\hline                 
Date & R.A. & DEC &  \# of fibers \\    
 & \multicolumn{2}{c}{Field centre (J2000)} & GIRAFFE/UVES/sky \\
\hline                        
2012-06-22 & 16:27:57.71 & -24:03:18.0 &  29/2/21\\
2012-06-22 & 16:26:57.60 & -24:05:48.0 &  28/3/24 \\
2012-06-24 & 16:25:40.03 & -23:58:27.4 &  32/1/25 \\
2012-06-24 & 16:29:00.09 & -24:28:14.0 &  21/5/22 \\
2012-06-25 & 16:25:18.07 & -24:40:04.0 &  37/1/28 \\
2012-06-25 & 16:25:55.93 & -24:25:07.0 & 16/1/20 \\
2012-06-25 & 16:29:11.49 & -24:55:05.1 &  29/4/22 \\
2012-06-25 & 16:25:40.43 & -24:57:25.9 & 36/4/30 \\
2012-08-30 & 16:26:56.61 & -24:53:13.2 &  15/4/15 \\
\hline                                   
\end{tabular}
\tablefoot{The last column reports the number of fibers allocated in each FLAMES field on GIRAFFE targets, UVES targets, and sky targets. }
\end{table}
%

\subsection{Data reduction and analysis} 

A summary of the data reduction process is reported in Sacco et al. (2014) and Lewis et al. (in prep.) for UVES and GIRAFFE spectra, respectively.  
A summary of the GIRAFFE data reduction is also reported in Jeffries et al. (2014). 
The detailed procedures adopted to retrieve the fundamental parameters (e.g., T$_{\rm eff}$, log$g$, $v$sin$i$), as well as the raw measurements of observed quantities (e.g. the equivalent widths of the H$\alpha$ line at 6562.8~\AA\ and the lithium line at 6707.8~\AA\ ($W(Li)$)) for pre-main sequence stars have been reported in Lanzafame et al. (2015).    

Throughout this manuscript we mainly use the recommended derived parameters for $W(Li)$, T$_{\rm eff}$, and radial velocity.
In the following sections, we briefly describe how each of these parameters was determined by the GES consortium and refer to specific papers for further details.

\subsubsection{Lithium equivalent width} 

The GES employed three independent methods to measure  $W(Li)$ in the GIRAFFE spectra: DAOSPEC (Stetson \& Pancino 2008), direct profile integration using the {\rm SPLOT} task within {\rm IRAF{\footnote{Image reduction and analysis facility}}}, and a semi-authomatic {\rm IDL{\footnote{Interactive Data Language}}} procedure developed for the GES. The latter two methods were also employed to measure $W(Li)$ in the UVES spectra. 
The final recommended $W(Li)$ values (actual detection or upper limit) are the average among the estimates derived by different methods, after discarding values that were inconsistent (with difference at a 20\% level).  We refer to Lanzafame et al. (2015) for a detailed explanation of the measurements of the lithium equivalent width. 
Out of 313 targets, the Gaia-ESO consortium recommended a $W(Li)$ value for 155 objects, an upper limit in this parameter for 146 objects, and no recommended $W(Li)$ for the remaining 12 objects. 

\subsubsection{Effective temperature  T$_{\rm eff}$}

T$_{\rm eff}$, together with other fundamental parameters such as log$g$ and $vsin~i$ are derived for all the objects with a signal-to-noise ratio (SNR) greater than 20. We again refer to  Lanzafame et al. (2015) for a detailed explanation of the method used to derive T$_{\rm eff}$. 
Among the 313 total objects analyzed by the GES team in the direction of L1688, 265 have an estimate of the effective temperature (244 GIRAFFE and 21 UVES targets). For the remaining 48 objects (46 GIRAFFE targets and two UVES targets), T$_{\rm eff}$ is not provided due to the low SNR, and we  discuss these objects at the end of the next section. 

\subsubsection{Radial velocities} 

Radial velocities for UVES targets were obtained following the techniques detailed in Sacco et al. (2014), while for GIRAFFE targets we refer to Jeffries et al. (2014), Jackson et al. (2015), and Koposov et al. (in prep.). 
Briefly, a cross-correlation method with a grid of synthetic spectra has been employed to give an initial estimate of the stellar radial velocity. Then a multi-parameter fit  of each spectrum with a template produced the adopted radial velocity value and corresponding uncertainty. 
The RV measurements are provided for all the 313 objects. In the following analysis we adopt the uncertainties on the RV measurements empirically determined using the prescription provided by Jackson et al. (2015), where they use the differences in RV measured between repeated observations to determine the underlying distribution of measurement uncertainties.  The RV determinations have a mean precision $\lesssim$0.27~km~s$^{-1}$. 
We note that for this sample of objects, the uncertainties provided by the GES and the one retrieved with this method are similar.

\section{Membership selection}
\label{memberSect}

One of the aims of the Gaia-ESO survey is to provide the astronomical community with reliable membership lists of stars in open clusters. 
The selection criteria we adopted to reach this goal imply that a large number of non-members are also observed.  In this section, we identify members among the objects analyzed by GES in the direction of L1688,  making use of both the spectroscopic information (equivalent widths of the lithium line) and the position of the stars in the Hertzsprung-Russell (HR) diagram. 

Absorption by photospheric lithium is a good proxy for youth in late-type stars. In fact, lithium rapidly burns once the base of the convection zone or the core temperature in fully convective stars reach $\sim3\times10^6$~K. The timescale for significant lithium depletion depends on the stellar mass (hence luminosity and temperature): M-type stars with lithium are younger than 10-20~Myr, K-type stars reach the lithium burning temperatures after $\sim$100~Myr, and G-type stars much later ($\sim$1~Gyr)  (see Soderblom 2010 for a review). The presence of Li excludes the vast majority of main sequence K- and M-dwarfs, while contamination by Li-rich field giants is still possible as only $\sim1-2$\% of G/K giants might show photospheric lithium (e.g., Brown et al. 1989; Smith et al. 1995, among many others). 

\begin{figure}
   \centering
      \includegraphics[angle=0,width=9cm]{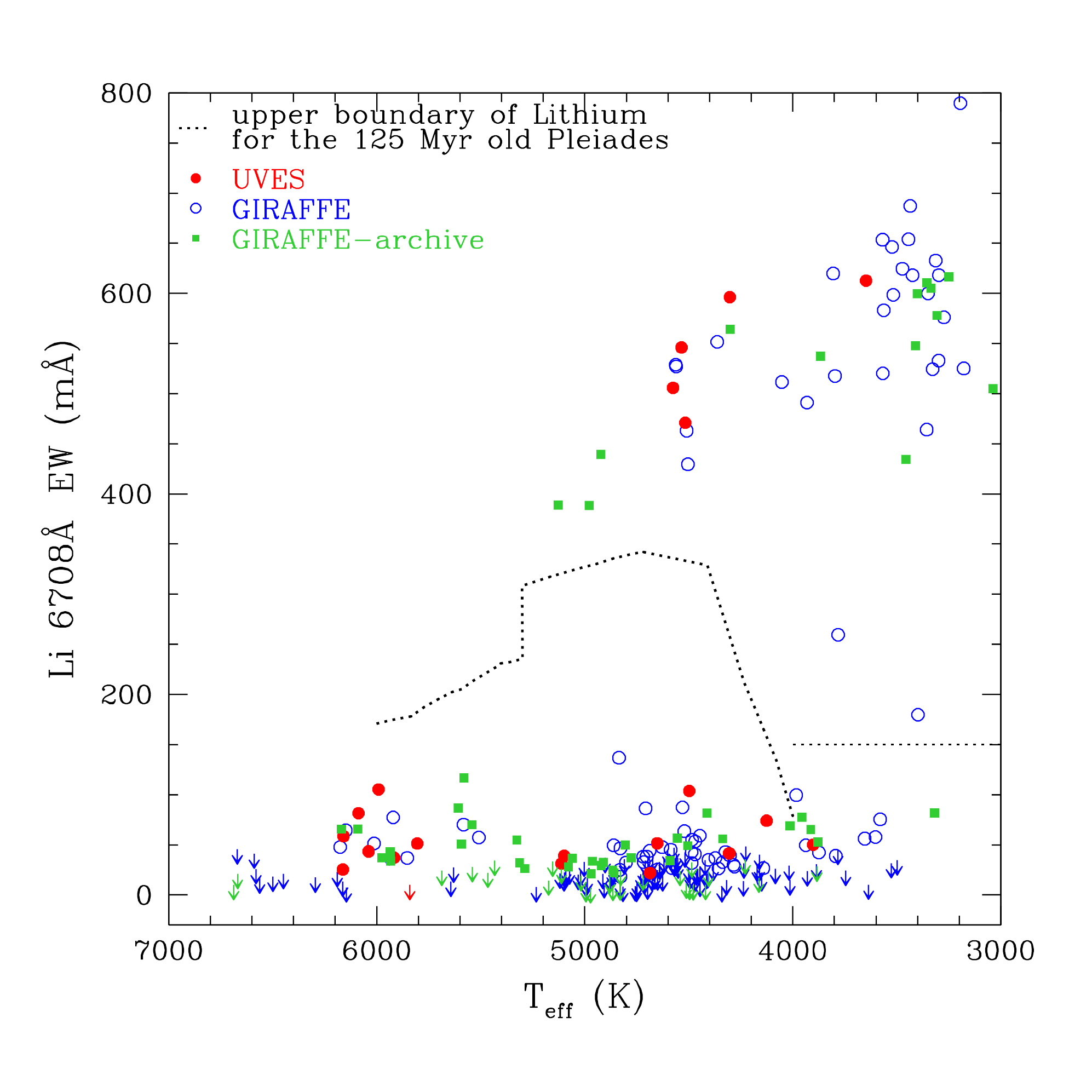}
   \caption{Lithium equivalent width ($W(Li)$) versus T$_{\rm eff}$. Symbols as in Fig.~1. The upper boundary of lithium depletion for the Pleiades is shown as a dotted line, together with the threshold at 150 m$\AA$ for stars with T$\rm_{eff}<$4000~K. }
              \label{Teff_vs_Li}%
    \end{figure}

Figure~\ref{Teff_vs_Li} shows $W(Li)$ as a function of the T$_{\rm eff}$ compared with the upper envelope of lithium depletion for the Pleiades (Stauffer et al. 1998). The latter is meant to represent the lithium equivalent width of stars as old as $\sim$125~Myr spanning a similar range of  T$_{\rm eff}$ as our sample.  Among the 265 objects with an estimate of  T$_{\rm eff}$ and $W(Li)$, 140 have a recommended value for $W(Li)$ while 125 have only an upper limit. 
The upper envelope of lithium for the Pleiades marks the threshold with which to discern candidate members. For stars with T$_{\rm eff}<$4000~K, we set the threshold to $\sim$150~\rm{m\AA}. Based on this criterion, we define 47 candidate members. 

\begin{figure}
   \centering
    \includegraphics[angle=-90,width=9.5cm]{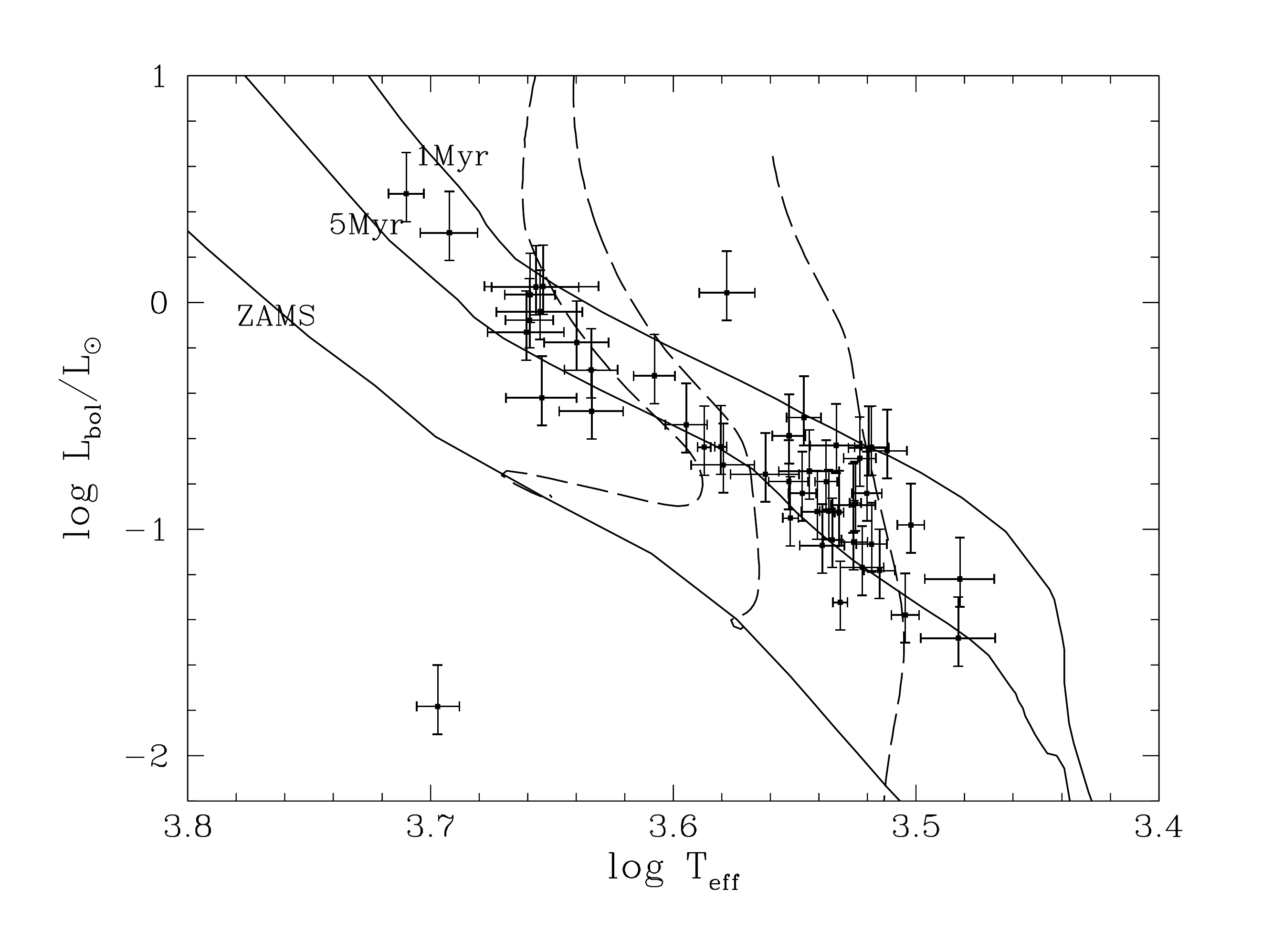}
   \caption{HR diagram of the members selected as candidates because of their lithium content. The evolutionary tracks for 0.2, 0.5 and 1.0~M$_{\odot}$ from D'Antona \& Mazzitelli (1997, 1998) are drawn as dashed lines. Solid lines show the isochrones and the zero-age main sequence, as labeled. }
              \label{FigHRD}%
    \end{figure}

These lithium-selected targets are then tested against their location on the HR diagram. The purpose is to exclude both lithium-rich field objects that exhibit photospheric lithium at a level that matches our threshold and objects lying below the Zero-Age Main Sequence (ZAMS) that are likely more distant than L1688. We computed bolometric luminosities from the $I$-band magnitudes, dereddening the observed magnitudes using the $R-I$ color excess and the Cohen et al. (2003) reddening law. The intrinsic colors and bolometric corrections were derived from Pecaut \&  Mamajek (2013). Following Wilking et al. (2005), the uncertainty on log$L$ is estimated to be 0.12~dex due to a combination of errors in $R$ and $I$ photometry, the distance, and the bolometric correction. 
Figure~\ref{FigHRD} shows the resulting HR diagram for the 47 objects analyzed within the GES.
Among the 47 stars selected through the $W(Li)$ thresholds, one located above the 1~Myr old isochrone is likely a Li-rich field giant and one lies well below the ZAMS. We therefore do not consider these objects in the following, basing our analysis on the remaining 45 objects, 28 and five observed with GIRAFFE and UVES, respectively, as part of the Gaia-ESO survey, and 12 collected from the ESO/GIRAFFE archive . 
The stellar properties for these 45 candidate members are reported in Table~\ref{Conf_members_table}.

Out of these 45 objects, 43 were previously known members from the literature (WMR05 and E11), and two objects (J16251469-2456069{\footnote {This object was identified as X-ray source by Martin et al. 1998.}} and J16244941-2459388) are identified here as new members. 

Finally, we consider the objects for which the GES does not provide all of the fundamental parameters (48 out of the 313 total), mainly due to low SNR ($<$20). 
 Four of these objects were previously proposed as candidate members by WMR05. We report in Table~\ref{tab_WMR_obj} these objects with the stellar parameters as derived by WMR05, and the radial velocity obtained by the GES. In the following analysis, we include the two objects with $W(Li)$ matching our lithium thresholds.

\begin{table*}
\caption{\label{Conf_members_table} GAIA/ESO candidate members of L1688.}
\centering
\begin{tabular}{ c c c c c c c c c c c c }
\hline\hline
CNAME & R.A. (J2000) & DEC (J2000) & L$_{star}$ & Mass & T$\rm_{eff}$ & RV &  $W(Li)$ & RV$_{gas}$ & Tracer\\
	&	hh:mm:ss & $^\circ:^\prime:^{\prime\prime}$ & L$_{\odot}$ &  M$_{\odot}$ & K & km~s$^{-1}$  & m\AA & & \\
\hline
16244941-2459388$^n$ & 16:24:49.41 & -24:59:38.8 & 0.05 & 0.28 & 3398 & -3.45 & 179.8 & ... &  ...\\  
 & & & & & (24) & (0.40) & (55.4) & & \\
16245974-2456008 & 16:24:59.74&-24:56:00.8 & 0.13 & 0.27 & 3356 & -6.01 & 464.1 & -5.18 & $^{12}$CO\\  
 & & & & & (71) & (0.29) & (7.8) & &\\  
16251469-2456069$^n$ & 16:25:14.69&-24:56:06.9 & 0.50 & 1.04 & 4303 & -2.53 & 596.3 & -5.34 & $^{12}$CO \\  
 & & & & & (109) & (0.48) & (40.3) & &\\  
16252243-2402057 & 16:25:22.43&-24:02:05.7 & 1.08 & 1.29 & 4560 & -7.54 & 527.2 & -7.37 & $^{13}$CO\\  
& & & & & (109) & (0.25) & (4.0) & &\\  
16252429-2415401 & 16:25:24.29&-24:15:40.1 & 0.04 & 0.16 & 3195 & -4.61 & 789.8 & -8.01 & $^{13}$CO$^*$ \\  
 & & & & & (42) & (1.82) & (94.0) & &\\ 
\hline
\end{tabular}
\tablefoot{A full version of the table is available as online material. The numbers in parenthesis represent the errors on the measured quantities. 
\tablefoottext{b}{Candidate binary system because of their measured RV.}
\tablefoottext{n}{New association members.}
\tablefoottext{*}{The gas tracer is self absorbed.}
}
\end{table*}

\begin{table*}
\caption{Confirmed members without stellar parameters recommended by GES.}             
\label{tab_WMR_obj}      
\centering                          
\begin{tabular}{ c c c c c c c c c c c} 
\hline\hline                 
CNAME & R.A. & DEC  & L$_{star}$\tablefootmark{A} & Mass\tablefootmark{A} & T$\rm_{eff}$\tablefootmark{A} & RV\tablefootmark{B} &  $W(Li)$\tablefootmark{B}  & Source Name\tablefootmark{C}\\
	&	hh:mm:ss & $^\circ:^\prime:^{\prime\prime}$ & L$_{\odot}$ &  M$_{\odot}$ & K & km~s$^{-1}$  & m\AA &  \\
\hline                        
16253958-2426349 & 16:25:39.58&-24:26:34.9 & 0.18 &  0.31 & 3499 & -9.27 (2.00) & 457.1 (20.0) & WLY 2-3 \\
16263416-2423282 & 16:26:34.16&-24:23:28.2 & 1580 &  5.00 & 18967 & -2.37 (2.71) & ... & Oph S1 \\
16282480-2435434 & 16:28:24.80&-24:35:43.4 & 0.03  & 0.14 & 2999 & ... & ... & [WMR2005](3-39) \\	
16284304-2422522 & 16:28:43.04&-24:22:52.2 & 0.06 &  0.15 & 3033 & -0.01 (3.53) & 757.6 (20.0) & [WMR2005](2-23) \\
\hline                                   
\end{tabular}
\tablefoot{The numbers in parenthesis represent the errors on the measured quantities. 
\tablefoottext{A}{Parameters from WMR05 (the values for the luminosity listed assumed a distance of 150 pc);}
\tablefoottext{B}{Parameters from GES;}
\tablefoottext{C}{Identification number as used by WMR05 in their Table~4.}
}
\end{table*}

In summary, the membership analysis based on the $W(Li)$ thresholds and location on the HR diagram gives a total of 47 objects as candidate members of the L1688 region around the $\rho$~Ophiuchi molecular cloud in a mass range between $\sim$0.2--1.7~M$_{\odot}$. 
Accurate RV values have been released by the GES consortium for these objects.

%

\section{Analysis and results}

\subsection{Radial velocity distribution of L1688} \label{sect_rv_distribution}

\begin{figure}
   \centering
    \includegraphics[angle=-90,width=9cm]{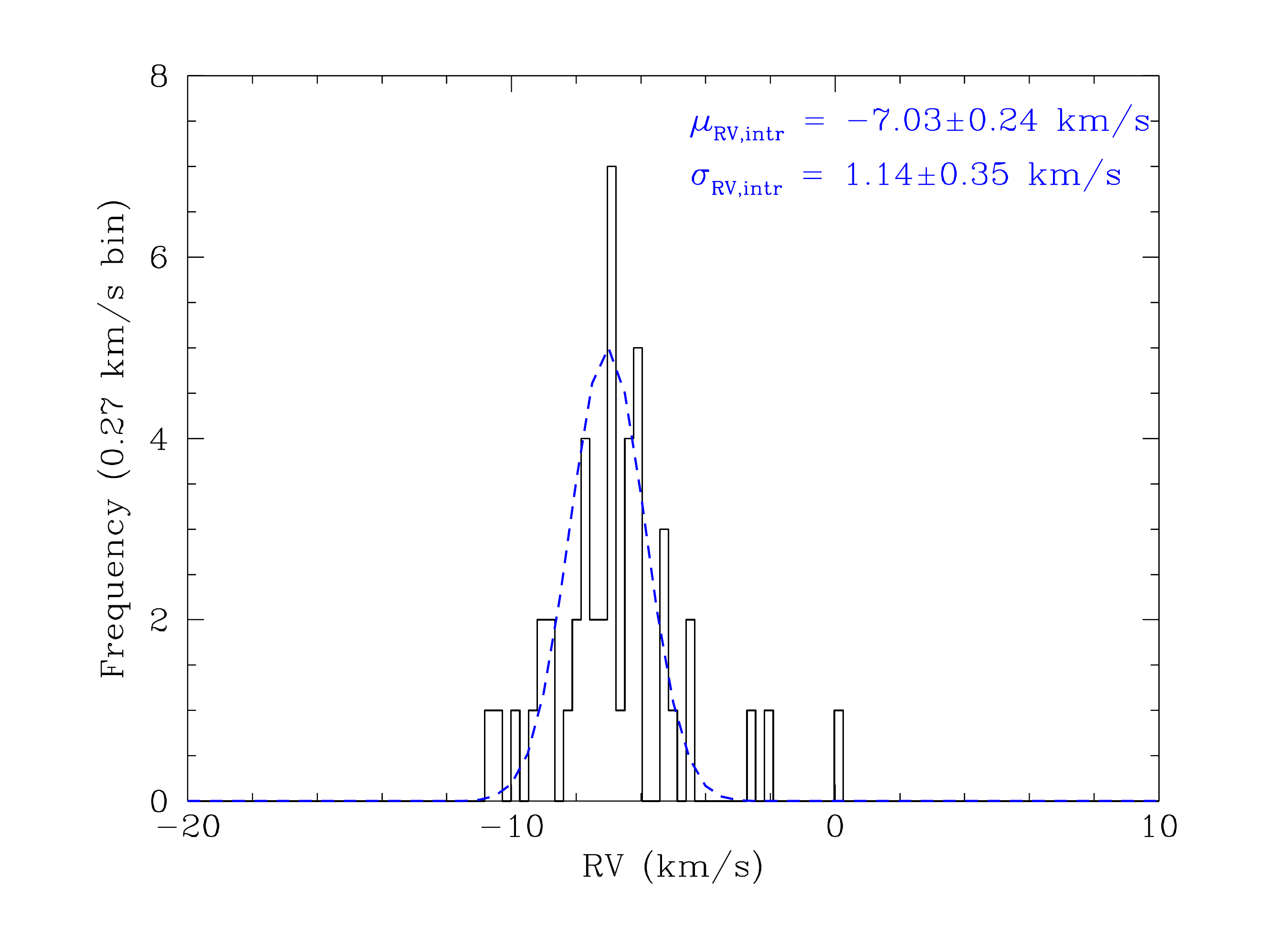}
   \caption{RV histogram for the L1688 candidate members. The data have been binned to 0.27~km~s$^{-1}$ that corresponds to the median RV uncertainty. The best fit model to the data accounting for unresolved binaries (with binary fraction 0.56) and uncertainties in the radial velocity measurements is shown as a blue dashed profile. }
              \label{FigRV}%
    \end{figure}

The observed distribution of the radial velocities is shown in Fig.~\ref{FigRV}. All but one object has a RV in the range $-20<RV<0$~km~s$^{-1}$. The only star outside this range is likely a close binary system, and it is not included in the following analysis.  

We modeled the observed radial velocity distribution using a maximum likelihood technique, as developed by Cottaar et al. (2012) and Cottaar \& H\'enault-Brunet (2014). In summary, they assumed that the observed RVs (as shown in Fig.~\ref{FigRV}) were drawn from an intrinsic distribution that is further broadened by unresolved binary orbital motions at large separations and by the uncertainty in the RV measurements. They assumed a universal companion mass ratio and period distribution appropriate for FGK stars which dominate our sample (Reggiani \& Meyer, 2013; Raghavan et al. 2010, respectively). The uncertainties in the parameters are computed by Markov Chain Monte Carlo (MCMC) simulations, as explained in Cottar et al. (2012) and Cottaar et al. (2015). 
The resulting {\it intrinsic} distribution, shown as a blue profile in Fig.~\ref{FigRV}, is centered at $\mu_{RV,intr}$=-7.03$\pm$0.24 km~s$^{-1}$ with a velocity dispersion of $\sigma_{RV,intr}$=1.14$\pm$0.35~km~s$^{-1}$, and a corresponding binarity fraction of $\sim$0.56. 
We also performed the fit where we kept the fraction of binaries fixed at 0.5. The mean velocity and dispersion of the cluster obtained in this case ($\mu_{RV,intr}$=-7.00$\pm$0.24 km~s$^{-1}$, $\sigma_{RV,intr}$=1.17$\pm$0.33~km~s$^{-1}$) were consistent with the results found when the binary fraction was left free to vary. 

\subsection{Stellar dynamics of L1688}

The dynamical state of L1688 can be understood by comparing our measurements of the RV dispersion ($\sigma_{RV,intr}$) with the velocity dispersion expected for a cluster in virial equilibrium ($\sigma_{vir}$) with mass $M_{dyn}$ and half-mass radius $r_{hm}$. 
$\sigma_{vir}$ is obtained by assuming that the region is gravitationally bound, with isotropic velocities and no mass segregation (Parker et al. 2012) and is defined by Binney \& Tremaine (1987) as: 
\begin{equation}
\sigma^2_{vir}=\frac{M_{dyn}~G}{\eta~r_{hm}}.
\label{EQ_sigmaRad}
\end{equation}

The total dynamical mass of the \rhoO molecular cloud is dominated by the total gas mass which has been estimated by Loren (1989) from $^{13}$CO emission lines to be $\sim$3050~M$_{\odot}$ (assuming a distance of 135 pc). Over half of it is concentrated around L1688. We estimated the dynamical mass $M_{dyn}$ of L1688 by summing the $^{13}$CO integrated intensity contained within the 1.3 deg$^2$ area centered on L1688 (the dashed box in Fig.~1 of E11). We computed $M_{dyn}$ by converting the $^{13}$CO integrated intensity to column density assuming T$_{ex}$=25~K and then to H$_2$ column density assuming H$_2$/$^{13}$CO$\sim$4.0$\times 10^5$ (Pineda et al. 2008). 
The uncertainties on $M_{dyn}$ are given by errors in the $^{13}$CO column density of 20-36\% in the different contour intervals, a 25\% error in the H$_2$/$^{13}$CO ratio, and a distance uncertainty of $\pm$10~pc.  Accounting for these uncertainties and assuming a mean molecular mass $\sim$2.33$\times$m$_H$, $M_{dyn}$=$\sim$1750$\pm$600~M$_{\odot}$ inside the dashed box.  
We have not considered in our analysis the total mass of the stars in L1688. The total mass of the 135 confirmed optically visible members analyzed by E11 is $\sim$85$M_{\odot}$. 
Consequently, the total mass in stars is probably less than $\sim$15\% of the error in $M_{dyn}$ and will be neglected in our calculation. 

The half mass radius $r_{hm}$ is approximated by fitting an ellipse to the molecular gas distribution containing half of the mass of the molecular gas. The projected mean half mass radius of the ellipse is $\sim$0.60~pc, with semi-major axis $a\sim$0.81~pc and semi-minor axis $b\sim$0.44~pc. To account for the 3-D structure of the region and the errors on the projected $r_{hm}$, we assumed an ellipsoid with a dimension in the $z$-direction equal either to the semi-minor axis or the semi-major axis of the projected ellipse.  As a result, we have estimated a mean radius of $r_{hm}$=0.64$\pm$0.09~pc.

The parameter $\eta$ is a numerical constant that depends on the density profile of the region (Portegies Zwart et al. 2010). The most widely used $\eta$-value for clusters ($\eta$=9.75) corresponds to the analytical result for a stellar volume density represented by a Plummer sphere (Plummer 1911).   Here $\eta=6\times r_{vir}/r{\rm_{eff}}$ \footnote{$r{\rm_{eff}}$ is defined as the projected half-light radius, see also Portegies Zwart et al. (2010). $r_{vir}$ is the virial radius defined as GM$^2$/2|U|, with M being the total mass of the cluster, and U the total potential energy.} and the surface density profile is given by $\Sigma(r)$=$\Sigma_0 \big(1+(\frac{r}{a})^2\big)^{-\gamma/2}$, with $a$ being a scale parameter and $\gamma$ the slope of the surface-density profile (Elson et al. 1987).   
In order to investigate the dynamical state of L1688, we also considered the upper and lower limits of the $\eta$ parameter. 
For $\gamma \lesssim$ 2 the Elson et al. (1987) profile has infinite mass, requiring $\eta >$6. The $\eta$ upper limit for this profile is obtained when $\gamma\sim$2.8 and corresponds to  $\eta \simeq$11 (see Fig.~4 in Portegies Zwart et al. 2010). 

\begin{figure}
   \centering
    \includegraphics[width=9cm]{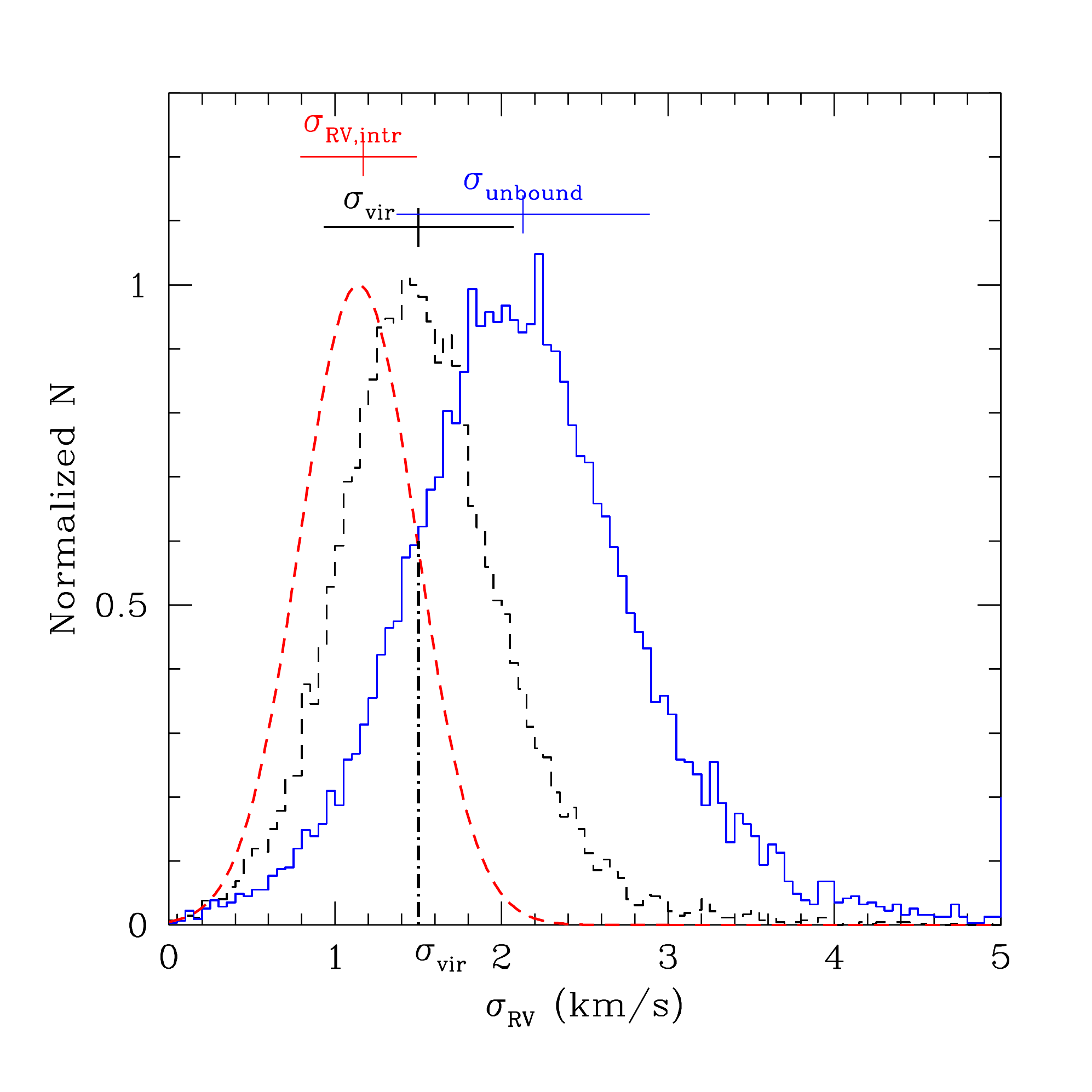}
   \caption{The black-dashed histogram represents the distribution of the velocity dispersions obtained from Eq.~(\ref{EQ_sigmaRad}) for 10,000 realizations, when errors on M$_{dyn}$ and r$_{hm}$, and the uncertainty on the $\eta$ parameter are taken into account. The corresponding distribution needed for the cluster to be unbound is shown as blue histogram. The distribution of the intrinsic radial velocity dispersion for L1688 is shown as red-dashed profile. 
}
              \label{Fig_comparison}%
    \end{figure}

Given the values of $M_{dyn}$, $r_{hm}$ and $\eta$ and their measurement errors, we computed $\sigma_{vir}$ and its associated error using a Monte Carlo approach. We calculated $\sigma_{vir}$ for 10,000 realizations, where for every realization we added a normally distributed error to $M_{dyn}$ and $r_{hm}$ and considered a random value for $\eta$ between 6 and 11. The final value for  $\sigma_{vir}$ was given by the mean and standard deviation of the 10,000 realizations and corresponded to $\sim$1.50$\pm$0.57~km~s$^{-1}$ (see Fig.~\ref{Fig_comparison}). 
Using the same approach we also constrained the radial velocity dispersion expected for the cluster if it were unbound ($\sigma_{unbound}$).  In this case the kinetic energy had to be bigger than the gravitational one, resulting in a velocity dispersion $\sqrt{2}$ larger than what is expected in virial equilibrium. Running 10,000 realizations, we found that the mean $\sigma_{unbound}$ was $\sim$2.13$\pm$0.76~km~s$^{-1}$.

We then employed a Bayesian analysis to measure the probability that the cluster is unbound (or bound) to the remnant local gas given the intrinsic radial velocity distribution.  Fig.~\ref{Fig_comparison} shows $\sigma_{RV,intr}$ (red gaussian profile) compared to the radial velocity distribution $\sigma_{vir}$ and $\sigma_{unbound}$ from the simulations, when the cluster is consistent with being in virial equilibrium or unbound, respectively.  
We defined $P(U|intr)$ as the probability that the cluster is unbound and $P(B|intr)$ as the probability that the cluster is bound given the observed intrinsic radial velocity distribution.  
The Bayesian evidence approach (e.g., Knuth et al. 2015) considers the Bayes factor (the odds ratio) given by: 
\begin{equation}
\frac{P(U|intr)}{P(B|intr)}=\frac{P(intr|U)P(U)}{P(intr|B)P(B)}
\end{equation}
where $P(U)$ and $P(B)$ are the prior probabilities the cluster in unbound or bound. We define $P(U)$=0.9 and $P(B)$=0.1, given that by an age of 5-10~Myr 90\% of the clusters are unbound (see references in the introduction). $P(intr|U)$ and $P(intr|B)$ are the probabilities to observe the intrinsic RV distribution given that the cluster is unbound or bound and was derived from the probability density function of the unbound gaussian distribution using the {\rm gauss\_pdf.pro} function in IDL. 
We found that the probability for L1688 to be currently unbound to the remnant local gas given the intrinsic radial velocity distribution is 0.20, and hence the probability for the cluster to be bound is 0.80. 
The prior used here are quite conservative. Recently, Kuhn et al. 2015 suggested that in star forming regions withh 500-10,000 stars before gas expulsion more than half of clusters are bound.

\subsection{Stellar velocity gradient}

\begin{figure}
   \centering\label{3d_plot_RV}
    \includegraphics[width=9cm, angle=0]{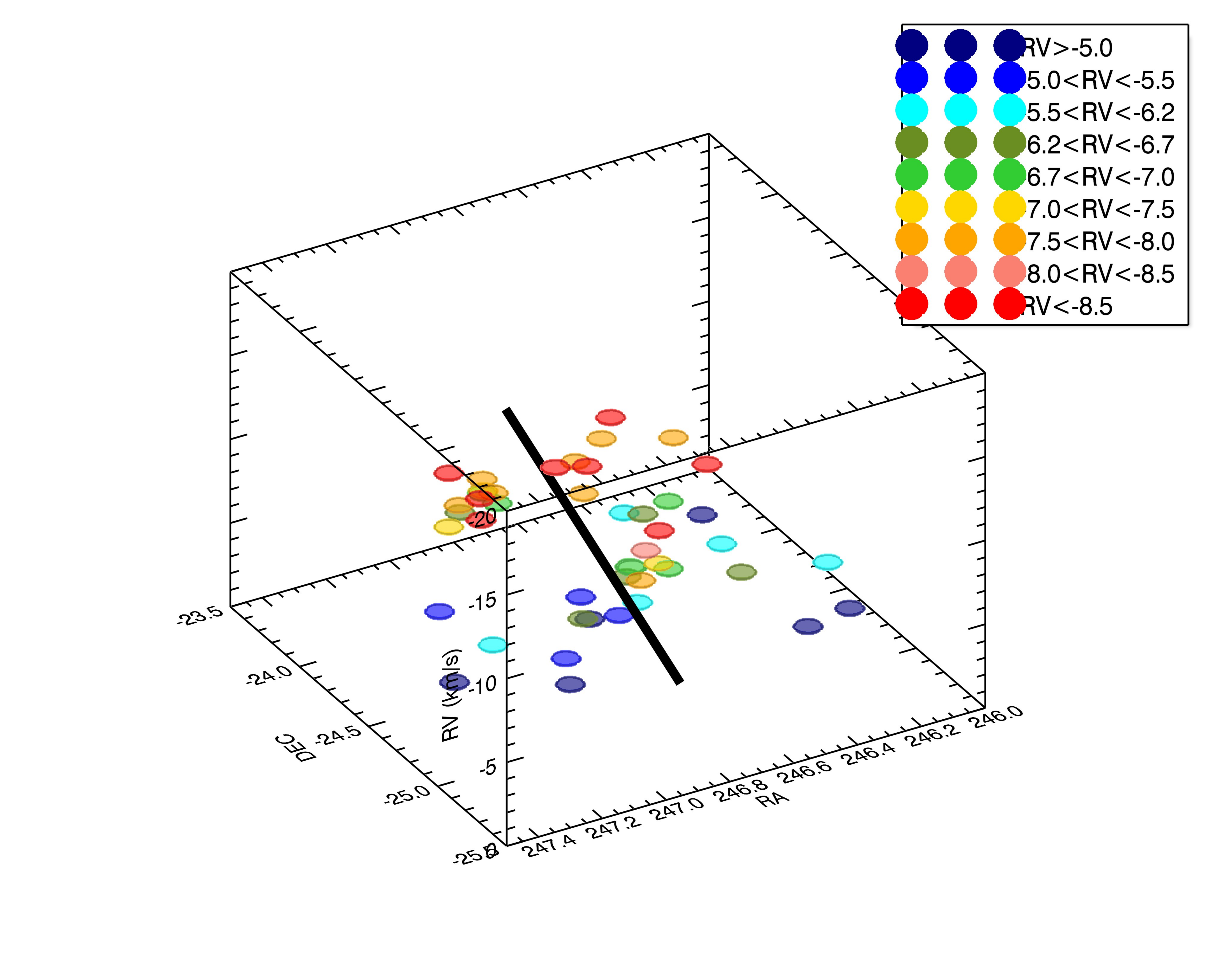}	
   \caption{Spatial distribution of the stellar surface population as a function of the radial velocity. The objects are divided in velocity bins. The black line represents the P.A.$\sim$160$^\circ$ direction of the gradient in the projected radial velocities.}
  \end{figure}

\begin{figure}
   \centering  \label{Fig_grad_vel}
    \includegraphics[width=7cm, angle=-90]{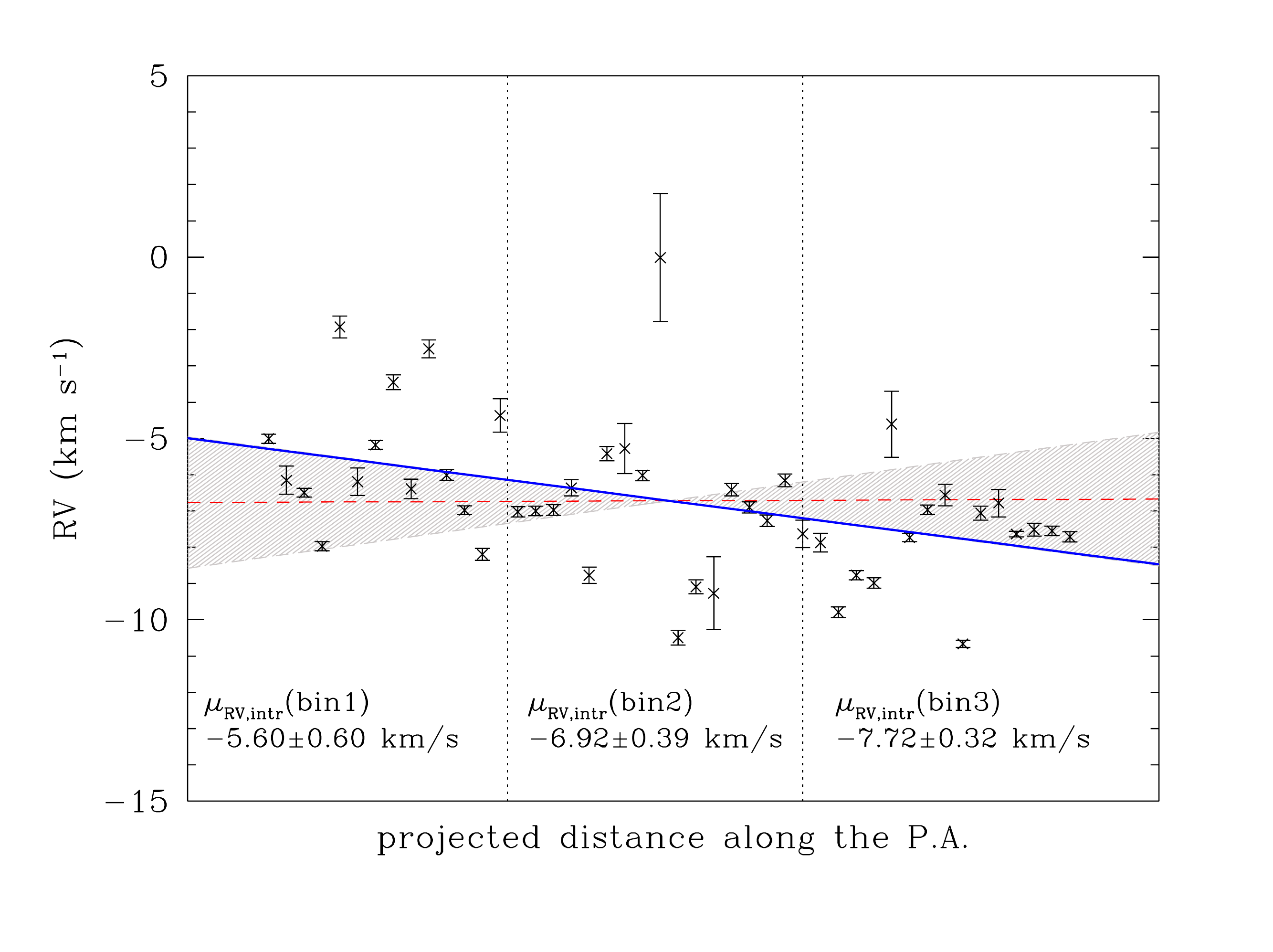}	
   \caption{Confirmed members as a function of the projected direction along the P.A.$\sim$160$^\circ$.  The blue-solid line represents the least squares fitting of the data. The red-dashed line and grey-shaded area are the mean of over 500 realizations where the data have been randomly distributed, and the 3$\sigma$ confidence level, respectively. The vertical-dotted lines divide the three bins we selected along the rotation angle. The mean radial velocity for each bin is reported.   }
  \end{figure}

To investigate further the dynamics of the L1688 cluster, we have checked for a gradient in the radial velocity of the stellar surface population. 
A 3-D representation of the stellar distribution is shown in Fig.~\ref{3d_plot_RV} where the z-dimension represents the radial velocity of the YSOs.   In order to find if there is a gradient, and because we do not know the distances of the YSOs, we have projected the stellar surface population onto the plane of the sky. 
We first divided the plane of the sky in two sections, centered on the median location of the YSOs' positions (RA=16:27:06.48 DEC=-24:22:40.8) and along a Position Angle (P.A.) = 0$^\circ$ (North-South).  We then rotated the P.A. counterclockwise, from 0$^\circ$ to 180$^\circ$ in steps of $\sim$9$^\circ$ (0.16 radians), corresponding to the minimum angle over which at least one object was included/removed into the following/previous section. For every angle we measured the difference between the mean intrinsic radial velocity of the two sections ($\Delta\mu_{RV}$). For P.A.$\sim$160$^{\circ}$, 
$\Delta\mu_{RV} = \mu_{RV,section1} - \mu_{RV,section2} \sim 0 \pm 0.6$~km~s$^{-1}$ meaning that along this direction the objects in the two different sections are moving on average with the same radial velocities.
We found that there is a gradient of increasing radial velocities of the members along this P.A. moving from the northwest to the southeast (this direction is identified by the black line in the 3-D representation in Fig.~\ref{3d_plot_RV}, and by the orange line in Fig.~\ref{FigFOV}).
To test whether this gradient is real, we randomly assigned the RV of each object to another object in another position, and repeated this operation 500 times.
Each time we measured the correlation between the RV and the projected distance (the gradient in RV).  Fig.~\ref{Fig_grad_vel} shows the mean (red line) and standard deviation (grey area) of the 500 realizations. 
With this test, we concluded that the confidence that the observed trend between RV and projected distance (blue line in Fig.~\ref{Fig_grad_vel}) is caused by a physical gradient is at the 3$\sigma$ level.  
To explore further this gradient, we divided the confirmed members into three bins along the rotation axis containing about the same number of objects (see  Fig.~\ref{Fig_grad_vel}). 
The intrinsic mean RV in each bin displays the same behavior, increasing as we move from the northwest to the southeast direction along the rotation axis, with a gradient in RV of $\sim$1.1~km~s$^{-1}$bin$^{-1}$. Since the area under consideration is $\sim$1.4~deg$^2$ in total, we find that the projected velocity gradient is $\sim$0.4~km~s$^{-1}$deg$^{-1}$, or $\sim$1.0~km~s$^{-1}$pc$^{-1}$ at a distance of 135~pc. 

A global gradient across the cloud has also been found in the past for dense gas tracers (N$_2$H$^+$ and DCO$^+$, Andr\'e et al. 2007, Pattle et al. 2015, Loren et al. 1990) and low-density gas tracers ($^{13}$CO, Loren 1988, Nutter et al. 2006).  In particular, for L1688 as traced by N$_2$H$^+$ (red box in Fig.~\ref{FigFOV}), Andr\'e et al. (2007) found a gradient of $\sim$1.1~km~s$^{-1}$~pc$^{-1}$ in the northwest to southeast direction, with a P.A.$\sim$120$^\circ$. This P.A. is consistent with the one found in this work through the radial velocity of the stellar surface population. 

In general, a gradient in the velocity of the members has been attributed to sequential and triggered star formation. Preibisch \& Zinnecker (1999) suggested a picture in which the massive stars in Upper-Centaurus-Lupus triggered the star formation in Upper Scorpius just as Upper Scorpius was triggering star formation in the $\rho$~Ophiuchi cloud complex.  
New age determinations of the low-mass members of the Upper Sco OB Association (Herczeg \& Hillenbrand 2015) estimate an age of $\sim$4~Myr. If the age of Upper Sco is 4~Myr, then there would have been enough time for a supernova event from a massive star formed at about the same age as the lower mass members to trigger the star formation in L1688.  Such an event $\sim$3 Myr ago would rule out the scenario proposed by Hoogerwerf et al. (2001) involving the runaway star $\zeta$ Oph in which the supernova explosion occurred only 1 Myr ago.  Alternatively, given the similarity in ages between the YSOs in this study and Upper Sco, a common triggering event from massive stars in Upper Centaurus Lupus is also plausible. 
Taking into account the $\sim$1.0 ~km~s$^{-1}$~pc$^{-1}$ (one dimensional deceleration) gradient in velocity, the shock wave produced by the supernova event could have had time to trigger the star formation in L1688 and initiated the formation of 2--3~Myr old stars. 
This scenario should be further investigated with detailed models in which the initial gas in L1688 is shocked and compressed to initiate the strong star formation activity.

The observed gradient in the mean radial velocity might also be attributed to the rotation of the cloud around the cluster center, with the southeastern side moving away from the observer, and the northwestern end moving toward us. We have associated the velocity of the rotation ($v_{rot}$) as the projected gradient velocity found previously in this section, in km~sec$^{-1}$~arcmin$^{-1}$. Considering this as a solid-body rotation, we defined $\mu_{RV,de-projected}$ as the radial velocity of each star around the cluster center when the rotation is taken into account: 
\begin{equation}
\mu_{RV,de-projected} = v_{rot} \, r \, cos(\theta-\alpha)+\mu_{RV,intr}
\end{equation}
where $r$ and $\theta$ are the polar coordinates of each star, and $\alpha$ is the angle of the velocity gradient. The values for $\mu_{RV,de-projected}$ were within the errors of $\mu_{RV,intr}$, and the slope of the trend shown in Fig.~\ref{Fig_grad_vel} is not affected. 
We then conclude that the trend of increasing radial velocities along the P.A.$\sim$160$^\circ$ from the northwest to the southeast is a real gradient.

\subsection{Stellar, dense-cores, and diffuse-gas velocity dispersions} 

The molecular gas in the cloud is well traced by $^{12}$CO and $^{13}$CO emission lines. Using the COMPLETE survey (Ridge et al. 2006), we determined the velocity of the local gas at each stellar position, as reported in Table~\ref{Conf_members_table}. We have found that there is no correlation between the stellar radial velocity and the diffuse gas velocity, as also found by Foster et al. (2015) for NGC~1333. 

We also compared the stellar radial velocity dispersion with the velocity dispersion of the dense cores{\footnote{We stress that we are always considering one-dimensional velocity dispersions.}}. Based on N$_2$H$^+$(1-0) observations, Andr\'e et al. (2007) analyzed the kinematics of starless condensations{\footnote{Starless cores are concentrations of molecular gas and dust without embedded stars. They can either evolve into pre-stellar cores (denser, self-gravitating and more centrally-concentrated) or they will eventually disperse and never form stars.}} in L1688, finding that they are gravitationally bound and pre-stellar in nature. 
Their data showed that very dense pre-stellar cores exhibit a subvirial velocity dispersion ($\sim$0.4 km~s$^{-1}$) relative to the mean of the ensemble.  A similar relationship between the velocity dispersion of the YSOs and dense cores has been observed in other star-forming regions (e.g., Myers 1983; Goodman et al. 1998; Caselli et al. 2002; Tafalla et al. 2004; Kirk et al. 2007; Andr{\'e} et al. 2007; Lada et al. 2008, Foster et al. 2015, Sacco et al. in prep.). 

The inconsistency between subvirial pre-stellar cores and virial or supervirial stars has been investigated in recent years.
One of the current ideas is that dense cores form in velocity-coherent filamentary clouds formed from converging turbulent flows (e.g., Elmegreen 2007; Gong \& Ostriker 2011).  It has been proposed that the higher velocity dispersion of the YSOs could arise due to magnetic fields constraining the dense cores (Foster et al. 2015), or the global collapse of the cluster that would convert gravitational potential energy into kinetic energy and thereby increase the probability of stellar encounters (Andr\'e et al. 2007; Foster et al. 2015).  
While it is not clear that magnetic fields have sufficient strength in L1688 to affect core dynamics (e.g., Troland et al. 1996), it is plausible that a global collapse is occurring, perhaps the result of an external trigger as discussed in Sect 4.3 (see also Andr\'e et al. 2007). Alternatively, if dense cores fragment and form multiple stars (e.g., Guszejnov \& Hopkins 2015) then the shorter crossing time for a typical core (10$^5$ years) could enable 2--3 Myr old YSOs to experience multiple stellar encounters that would pump up their velocity dispersion.  This last scenario would, however, be more valid if the spatial distribution of YSOs and dense cores were co-spatial, as in NGC~1333 (Foster et al. 2015).  
A similar interpretation has been previously proposed by Bate et al. (2003), who suggested that a difference between the velocity dispersions of dense cores and stars might be due to dynamical interactions between young stars that will disperse (or eject) them in random directions from the core where they formed. 
In particular, Bate et al. (2003) analyzed the collapse of a 50~M$_{\odot}$ turbulent molecular cloud with a free-fall time ($t_{\rm ff}$) of 1.90$\times$10$^5$~yr. After an initial period of chaotic interactions and ejections ($\sim$1.40~$t_{\rm ff}$), the velocity dispersion settled to a 1-D value of $\sim$1.2~km~s$^{-1}$, roughly a factor of three greater than the initial velocity dispersion of the gas.  They also compared their simulation with dense cores in L1688, finding that it produces dense cores comparable to those in L1688.  Later, Bate (2009) also investigated the kinematic structure of the gas of a collapsing cloud with a steeper spectrum for the initial turbulent velocity field, finding a slightly bigger velocity dispersion ($\sim$1.7~km s$^{-1}$), but indistinguishable statistical properties of the formed stars. Their result suggests that the evolution of the stellar population is mainly due to the effects of competitive accretion and dynamical interaction and ejection, and is almost non-dependent on the initial kinematic structure of the gas. 

We have estimated the crossing and relaxation timescales for the stellar surface population observed with GES.  Given the intrinsic velocity dispersion of $\sim$1.14~km~s$^{-1}$ we have found that the crossing time is $t_{cross,stars}$=R$_{cluster}$/$\sigma_{1D}$=0.9$\times$10$^6$~yr. The relaxation timescale  is defined  as $t_{relax,stars}=t_{cross,stars}\times N/(8ln(N/2))$ where N is the number of stars (Binney \& Tremaine 1987). 
The relaxation time is estimated to be  $t_{relax,stars}$=3.1$\times$10$^6$yr, where N=110. 
The ratio between the crossing time and the age of the cluster ($t_{cross}/age$) can be used to separate bound clusters from unbound associations (Gieles \& Portegies Zwart 2011). For L1688,  $t_{cross}$/age$\lesssim$1 suggesting the cluster is gravitationally bound, albeit through the binding mass of the molecular gas. 

Finally, we have investigated the impact of drastic gas expulsion in L1688. Several theoretical investigations have been made in the past years showing that if the gas is lost instantaneously and the star formation efficiency (SFE) is below $\sim$35\%, then the entire cluster should be disrupted with all the stars leaving the cluster's potential well and dispersing into the field (e.g., Hills  1980;  Lada et al. 1984,  Kroupa, Petr \& McCaughrean 1999; Bastian \& Goodwin 2006, Baumgardt \&  Kroupa 2007, Dale et al. 2015, among many others).  We have made the same calculation for L1688, assuming a SFE$\sim$10\% (Jorgensen et al. 2008). An instantaneous drastic removal of gas will leave the stars in a supervirial state, causing them to expand and disperse into the Galactic field.

\section{Summary}

We have carried out a spectroscopic study of the dynamical properties of the L1688 star forming region in the $\rho$~Ophiuchi cloud complex using the dataset provided by the Gaia-ESO survey.  The main findings of this work can be summarized as follows: 
\begin{itemize}
\item Membership selection of the L1688 cloud has been made based on the presence of the lithium absorption line in the spectra and by the location of the stars in the HR diagram.  A total of 47 objects were identified, with 45 already known as members (WMR05 and E11) and two as new candidate members.  Given the high extinction of the central region of L1688, the observed members can be considered representative of an older surface population of the young cluster. 
\item The radial velocity distribution of the YSOs is well represented by a single Gaussian. An intrinsic radial velocity dispersion of $\sigma_{RV,intr}$=1.14$\pm$0.35~km~s$^{-1}$ is obtained after accounting for undetected binaries and errors in the RV measurements. 
\item The velocity dispersion of the L1688 cluster is consistent with virial equilibrium, and the cluster is currently bound to the remnant gas with a $\sim$80\% confidence level.  
\item A gradient in the radial velocities of the stellar population of $\sim$1.0~km~s$^{-1}$pc$^{-1}$ has been identified along a northwest/southeast direction . This gradient may be related to the triggering of star formation by a supernova explosion in the Sco-Cen OB association. We have excluded the possibility that the observed gradient is due to cloud rotation.  
\item We have compared the stellar radial velocities to the velocity of the molecular gas in the cluster, traced by $^{12}$CO and $^{13}$CO emission lines, and with the velocity dispersion of the dense cores.  Pre-stellar dense cores exhibit a subvirial velocity dispersion that is a factor of three smaller than the stellar velocity dispersion. Despite the wealth of information obtained through accurate radial velocity measurements in L1688, the reason why dense gas cores have lower velocity dispersions compared to the YSO surface population is still not totally understood, however, a likely explanation is the dynamical interactions between YSOs as proposed by Bate et al. (2003). 
 \end{itemize}

By the end of the Gaia-ESO survey, when observations for about 30 young clusters will be completed, a comparison between stellar clusters with precise information on the stellar radial velocities and the velocity dispersions of dense pre-stellar cores can be made. The goals of this comparison are to gain a broader understanding of the fraction of star clusters that remain bound after gas dispersal rather than dissolving into the field, and how cluster properties (such as density, mass segregation, and mass) might affect their future behavior. 

\begin{acknowledgements} 
We thank the anonymous referee for useful suggestion that improved the clarity of the paper. 
E.R. and M.M. acknowledges financial support from the Swiss National Science Foundation (n. 200020-144492)
F.J.E. acknowledges financial support from the ARCHES project (7th Framework of the European Union, n. 
313146). 
A. Bayo acknowledges financial support from the Proyecto Fondecyt de Iniciaci\'on 11140572.
Based on data products from observations made with ESO Telescopes at the La Silla Paranal Observatory under programme ID 188.B-3002. These data products have been processed by the Cambridge Astronomy Survey Unit (CASU) at the Institute of Astronomy, University of Cambridge, and by the FLAMES/UVES reduction team at INAF/Osservatorio Astrofisico di Arcetri. These data have been obtained from the Gaia-ESO Survey Data Archive, prepared and hosted by the Wide Field Astronomy Unit, Institute for Astronomy, University of Edinburgh, which is funded by the UK Science and Technology Facilities Council.
This work was partly supported by the European Union FP7 programme through ERC grant number 320360 and by the Leverhulme Trust through grant RPG-2012-541. 
We acknowledge the support from INAF and Ministero dell’ Istruzione, dell’ Universita’ e della Ricerca (MIUR) in the form of the grant "Premiale VLT 2012" and the grant "The Chemical and Dynamical Evolution of the Milky Way and Local Group Galaxies" (prot. 2010LY5N2T). The results presented here benefit from discussions held during the Gaia-ESO workshops and conferences supported by the ESF (European Science Foundation) through the GREAT Research Network Programme.
\end{acknowledgements}

\onllongtab{
\begin{longtab}
\centering
\begin{longtable}{ c c c c c c c c c c c c }
\caption{\label{Conf_members_table_online} GAIA/ESO candidate members of L1688.}\\
\hline\hline
CNAME & R.A. (J2000) & DEC (J2000)  & L$_{star}$ & Mass & T$\rm_{eff}$ & RV &  $W(Li)$ & RV$_{gas}$ & Tracer \\
	&	hh:mm:ss & $^\circ:^\prime:^{\prime\prime}$ & L$_{\odot}$ &  M$_{\odot}$ & K & km~s$^{-1}$  & m\AA & & \\
\hline
\endfirsthead
\caption{continued.}\\
\hline\hline
CNAME & R.A. & DEC  & L$_{star}$ & Mass & T$_{eff}$ & RV &  $W(Li)$ & RV$_{gas}$ & Tracer \\
	&	hh:mm:ss & $^\circ:^\prime:^{\prime\prime}$ & L$_{\odot}$ &  M$_{\odot}$ & K & km~s$^{-1}$  & m\AA & & \\
\hline
\endhead
\hline
\endfoot
16244941-2459388\tablefootmark{n} & 16:24:49.41 & -24:59:38.8 & 0.05 & 0.28 & 3398 & -3.45 & 179.8 & -- &  --\\  
 & & & & & (24) & (0.40) & (55.4) & & \\
16245974-2456008 & 16:24:59.74&-24:56:00.8 & 0.13 & 0.27 & 3356 & -6.01 & 464.1 & -5.18 & $^{12}$CO\\  
 & & & & & (71) & (0.29) & (7.8) & &\\  
16251469-2456069\tablefootmark{n} & 16:25:14.69&-24:56:06.9 & 0.50 & 1.04 & 4303 & -2.53 & 596.3 & -5.34 & $^{12}$CO \\  
 & & & & & (109) & (0.48) & (40.3) & &\\  
16252243-2402057 & 16:25:22.43&-24:02:05.7 & 1.08 & 1.29 & 4560 & -7.54 & 527.2 & -7.37 & $^{13}$CO\\  
& & & & & (109) & (0.25) & (4.0) & &\\  
16252429-2415401 & 16:25:24.29&-24:15:40.1 & 0.04 & 0.16 & 3195 & -4.61 & 789.8 & -8.01 & $^{13}$CO$^*$ \\  
 & & & & & (42) & (1.82) & (94.0) & &\\ 
16254767-2437394 & 16:25:47.67&-24:37:39.4 & 0.12 & 0.32 & 3473 & -6.02 & 624.7 & -7.36 & $^{12}$CO\\  
 & & & & & (52) & (0.29) & (8.1) & &\\ 
16255893-2452483 & 16:25:58.93&-24:52:48.3 & 0.09 & 0.23 & 3298 & -6.39 & 618.2 & -6.46 & $^{12}$CO \\  
 & & & & & (48) & (0.54) & (14.6) & &\\
16255965-2421223 & 16:25:59.65&-24:21:22.3 & 0.23 & 0.27 & 3299 & -6.90 & 533.1 & -7.80 & $^{12}$CO\\ 
 & & & & & (52) & (0.31) & (5.2) & &\\ 
16260544-2355408 & 16:26:05.44&-23:55:40.8 & 0.14 & 0.26 & 3313 & -7.71 & 632.8 & -7.03 & $^{12}$CO\\ 
 & & & & & (47) & (0.27) & (25.0) & &\\ 
16261706-2420216 & 16:26:17.06&-24:20:21.6 & 0.74 & 1.19 & 4576 & -6.41 & 505.9 & -6.60 & $^{12}$CO\\
 & & & & & (166) & (0.35) & (24.6) & &\\  
16261877-2407190 & 16:26:18.77&-24:07:19.0 & 0.31 & 0.37 & 3518 & -10.66 & 598.5 & -6.67 & $^{13}$CO \\
 & & & & & (57) & (0.19) & (25.1) & &\\  
16262407-2416134 & 16:26:24.07&-24:16:13.4 & 1.18 & 1.30 & 4504 & -6.15 & 429.6 & -6.87 & $^{12}$CO\\
 & & & & & (244) & (0.35) & (18.6) & &\\  
16263297-2400168 & 16:26:32.97&-24:00:16.8 & 0.07 & 0.23 & 3328 & -7.51 & 524.5 & -6.86 & $^{12}$CO\\
 & & & & & (68) & (0.36) & (27.4) & &\\  
16264310-2411095 & 16:26:43.10&-24:11:09.5 & 0.29 & 0.68 & 3932 & -8.77 & 491.1 & -6.62 & $^{12}$CO\\
 & & & & & (78) & (0.26) & (4.1) & &\\  
16264429-2443141 & 16:26:44.29&-24:43:14.1 & 0.16 & 0.32 & 3444 & -8.77 & 654.0 & -6.70 & $^{13}$CO\\
 & & & & & (37) & (0.44) & (6.8) & &\\  
16264441-2447138 & 16:26:44.41&-24:47:13.8 & 0.09 & 0.26 & 3355 & -6.97 & 610.4 & -6.52 & $^{12}$CO\\
 & & & & & (45) & (0.25) & (4.9) & &\\  
16264705-2444298 & 16:26:47.05&-24:44:29.8 & 0.12 & 0.29 & 3402 & -7.02 & 599.7 & -6.66 & $^{13}$CO\\ 
 & & & & & (14) & (0.29) & (8.9) & &\\  
16264864-2356341$^b$ & 16:26:48.64&-23:56:34.1 & 0.47 & 0.79 & 4053 & -78.57 & 511.6 & -6.85 & $^{12}$CO\\
 & & & & & (79) & (0.26) & (3.3) & &\\  
16265048-2413522 & 16:26:50.48&-24:13:52.2 & 0.09 & 0.29 & 3424 & -7.63 & 618.1 & -6.61 & $^{13}$CO\\
 & & & & & (30) & (0.76) & (24.9) & &\\  
16265850-2445368 & 16:26:58.50&-24:45:36.8 & 3.01 & 1.72 & 5128 & -8.19 & 388.8 & -6.72 & $^{13}$CO\\
 & & & & & (86) & (0.33) & (2.7) & &\\  
16270405-2409318 & 16:27:04.05&-24:09:31.8 & 0.19 & 0.53 & 3797 & -8.98 & 517.8 & -6.83 & $^{13}$CO\\
 & & & & & (114) & (0.28) & (11.1) & &\\  
16270451-2442596 & 16:27:04.51&-24:42:59.6 & 0.33 & 0.90 & 4301 & -6.99 & 564.3 & -6.65 & $^{13}$CO\\ 
 & & & & & (130) & (0.27) & (12.7) & &\\ 
16270456-2442140 & 16:27:04.56&-24:42:14.0 & 0.23 & 0.60 & 3866 & -6.97 & 537.5 & -6.53 & $^{13}$CO \\ 
 & & & & & (25) & (0.30) & (9.2) & &\\ 
16270659-2441488 & 16:27:06.59&-24:41:48.8 & 0.03 & 0.08 & 3038 & -6.36 & 504.9 & -6.58 & $^{13}$CO\\  
 & & & & & (107) & (0.45) & (30.5) & &\\ 
16271513-2451388 & 16:27:15.13&-24:51:38.8 & 0.17 & 0.42 & 3648 & -6.19 & 612.8 & -6.41 & $^{12}$CO\\  
 & & & & & (118) & (0.75) & (30.4) & &\\
16271836-2454537 & 16:27:18.36&-24:54:53.7 & 0.08 & 0.30 & 3456 & -7.97 & 434.3 & -6.43 & $^{12}$CO\\  
 & & & & & (73) & (0.26) & (13.4) & &\\
16272297-2448071 & 16:27:22.97&-24:48:07.1 & 0.23 & 0.31 & 3411 & -5.18 & 547.8 & -6.56 & $^{12}$CO\\  
 & & & & & (83) & (0.24) & (4.8) & &\\
16273311-2441152 & 16:27:33.11&-24:41:15.2 & 2.03 & 1.57 & 4923 & -4.36 & 439.6 & -6.80 & $^{13}$CO\\  
 & & & & & (134) & (0.92) & (14.8) & &\\
16273526-2438334 & 16:27:35.26&-24:38:33.4 & 0.21 & 0.28 & 3335 & -5.42 & 605.1 & -6.80 & $^{13}$CO$^*$\\  
 & & & & & (51) & (0.39) & (12.1) & &\\ 
16273797-2357238 & 16:27:37.97&-23:57:23.8 & 0.16 & 0.38 & 3568 & -6.78 & 653.6 & -6.55 & $^{12}$CO\\  
 & & & & & (67) & (0.75) & (14.4) & &\\ 
16273832-2357324 & 16:27:38.32&-23:57:32.4 & 1.17 & 1.30 & 4535 & -7.63 & 546.1 & -6.55 & $^{12}$CO\\  
 & & & & & (187) & (0.16) & (32.7) & &\\ 
16273833-2404013 & 16:27:38.33&-24:04:01.3 & 0.67 & 1.09 & 4363 & -6.97 & 551.6 & -6.54 & $^{12}$CO\\  
 & & & & & (133) & (0.27) & (6.4) & &\\ 
16273901-2358187 & 16:27:39.01&-23:58:18.7 & 0.91 & 1.24 & 4517 & -7.06 & 471.0 & -6.67 & $^{12}$CO\\  
 & & & & & (184) & (0.39) & (28.3) & &\\
16274187-2404272 & 16:27:41.87&-24:04:27.2 & 0.26 & 0.39 & 3567 & -7.73 & 520.2 & -6.85 & $^{12}$CO\\  
 & & & & & (56) & (0.22) & (18.7) & &\\ 
16275996-2448193 & 16:27:59.96&-24:48:19.3 & 0.22 & 0.25 & 3250 & -1.93 & 616.4 & -6.62 & $^{12}$CO$^*$\\  
 & & & & & (61) & (0.61) & (29.6) & &\\
16280011-2453427 & 16:28:00.11&-24:53:42.7 & 0.23 & 0.27 & 3307 & -6.49 & 578.0 & -6.23 & $^{12}$CO\\  
 & & & & & (64) & (0.25) & (17.6) & &\\ 
16280080-2400517 & 16:28:00.80&-24:00:51.7 & 0.12 & 0.30 & 3435 & -6.56 & 687.2 & -6.55 & $^{13}$CO\\  
 & & & & & (30) & (0.59) & (10.2) & &\\ 
16281099-2406177 & 16:28:10.99&-24:06:17.7 & 0.13 & 0.27 & 3350 & -7.87 & 600.1 & -6.46 & $^{13}$CO\\  
 & & & & & (18) & (0.52) & (14.0) & &\\ 
16281673-2405142 & 16:28:16.73&-24:05:14.2 & 0.84 & 1.20 & 4563 & -9.79 & 528.9 & -6.53 & $^{13}$CO\\ 
 & & & & & (103) & (0.29) & (6.9) & &\\
16281922-2457340 & 16:28:19.22&-24:57:34.0 & 0.14 & 0.35 & 3523 & -5.01 & 646.3 & -- &  --\\  
 & & & & & (47) & (0.24) & (13.5) & &\\ 
16282151-2421549 & 16:28:21.51&-24:21:54.9 & 0.11 & 0.36 & 3563 & -9.09 & 583.3 & -6.44 & $^{13}$CO$^*$\\  
 & & & & & (26) & (0.38) & (16.7) & &\\
16282333-2422405 & 16:28:23.33&-24:22:40.5 & 0.38 & 0.92 & 4510 & -10.49 & 463.1 & -6.45 & $^{13}$CO$^*$\\  
 & & & & & (151) & (0.41) & (9.6) & &\\
16282430-2409316 & 16:28:24.30&-24:09:31.6 & 0.23 & 0.55 & 3806 & -7.27 & 619.8 & -6.58 & $^{13}$CO\\  
 & & & & & (22) & (0.32) & (32.9) & &\\
16290288-2427494 & 16:29:02.88&-24:27:49.4 & 0.10 & 0.19 & 3178 & -5.28 & 525.1 & -6.17 & $^{13}$CO\\ 
 & & & & & (41) & (1.38) & (17.3) & &\\
16290392-2451414 & 16:29:03.92&-24:51:41.4 & 0.07 & 0.21 & 3274 & -6.15 & 576.1 & -6.22 & $^{13}$CO\\  
 & & & & & (48) & (0.77) & (14.8) & &\\
\hline
\end{longtable}
\tablefoot{The numbers in parenthesis represent the errors on the measured quantities. 
\tablefoottext{b}{Candidate binary system because of their measured RV.}
\tablefoottext{n}{New association members.}
\tablefoottext{*}{The gas tracer is self absorbed.}
}
\end{longtab}
}


\end{document}